\documentclass[lineno]{jfm}

\usepackage{graphicx}
\usepackage{multirow}
\usepackage{amssymb}
\usepackage{amsmath}
\usepackage{bm}
\usepackage{newtxtext}
\usepackage{newtxmath}
\usepackage{natbib}
\usepackage[normalem]{ulem}
\usepackage{float}
\usepackage{hyperref}
\hypersetup{
    colorlinks = true,
    urlcolor   = blue,
    citecolor  = black,
}
\newcommand{\RomanNumeralCaps}[1]
\linenumbers
\usepackage{longtable}
\makeatletter

\newcommand{\Rmnum}[1]{\expandafter\@slowromancap\romannumeral #1@}
\makeatother

\title{Gravity drives the flow within the Stewartson layer in centrifugal convection}

\author{Rushi Lai\aff{1},
  Jun Zhong\aff{2},
 \and Chao Sun\aff{1,2}
 \corresp{\email{chaosun@tsinghua.edu.cn}}}

\affiliation{
\aff{1} Department of Engineering Mechanics, School of Aerospace Engineering, Tsinghua University, 100084 Beijing, China
\aff{2} New Cornerstone Science Laboratory, Center for Combustion Energy, Key Laboratory for Thermal Science and Power Engineering of Ministry of Education, Department of Energy and Power Engineering, Tsinghua University, 100084 Beijing, China}

\begin{document}
\maketitle

\begin{abstract}
We conduct three-dimensional numerical simulations on centrifugal convection (CC) in a closed annular container, incorporating gravity and no-slip top and bottom boundaries, to systematically investigate rotation-induced secondary flow. The Stewartson layer, identified by an elongated circulation in mean vertical velocity plots, emerges near the inner and outer cylinders only beyond a critical gravitational forcing.
Quantitative analyses confirm that the layer thickness scales as $\delta_{st}\sim Ek^{1/3}$ due to rotational effects, consistent with results from rotating Rayleigh-B\'enard convection, where $Ek$ represents the Ekman number. The internal circulation strength, however, is determined by both gravitational buoyancy and rotational effects. We propose that gravitational buoyancy drives the internal flow, which balances against viscous forces to establish a terminal velocity.
Through theoretical analysis, the vertical velocity amplitude follows $W_{st}\sim Ek^{5/3}Ro^{-1}Ra_gPr^{-1}$, showing good agreement with simulation results across a wide parameter range. Here, $Ro^{-1}$ represents the inverse Rossby number, $Ra_g$ the gravitational Rayleigh number, and $Pr$ the Prandtl number. The theoretical predictions match simulations well, demonstrating that the Stewartson layer is gravity-induced and rotationally constrained through geostrophic balance in the CC system.
These findings yield fundamental insights into turbulent flow structures and heat transfer mechanisms in the CC system, offering both theoretical advances and practical engineering applications.
\end{abstract}


\section{Introduction}


Convection-driven flows are ubiquitous in both natural environments and industrial applications \citep{2001_Tropical_Hartmann, 2000_Turbulent_Niemela, 1999_Open_Marshall, 2007_Rayleigh_King, 2014_AReview_Bairi, 2015_Review_Owen}. Rayleigh-B\'enard convection (RBC) serves as the classical paradigm of thermal convection, with a cooled upper plate and heated lower plate. The canonical planar RBC system has been extensively studied through experiments and simulations in recent decades \citep{2000_Scaling_Grossmann, 2004_Fluctuations_Grossman, 2009_Heat_Ahlers, 2010_Small_Lohse, 2012_New_F, 2024_Ultimate_Lohse}.
The investigation of RBC's ultimate regime at ultra-high Rayleigh numbers has been significantly advanced by centrifugal convection (CC), in which the exceptionally strong centrifugal forces enable unique experimental access to this challenging regime \citep{2020_Supergravitational_Jiang, 2022_Experimental_Jiang, 2021_Coriolis_Rouhi, 2022_Effects_Wang, 2023_From_Zhong}. The experimental results \citep{2022_Experimental_Jiang} from CC system confirm the predicted heat transfer scaling in the ultimate regime \citep{Kraichnan1962,2000_Scaling_Grossmann, 2024_Ultimate_Lohse}, with validation spanning more than an order of magnitude in Rayleigh number.
Most remarkably, CC reaches the ultimate regime at a critical Rayleigh number ($Ra_t \approx 10^{11}$) two to three orders of magnitude lower than planar RBC. This fundamental shift in the transition Rayleigh number requires thorough investigation of its physical origins.

In contrast to classical RBC, CC imposes centrifugal acceleration to enhance buoyant driving while introducing two additional fundamental factors: rotational effects and the gravity perpendicular to the temperature gradient. In practical experimental configurations, both factors may influence flow structures and heat transfer characteristics. However, previous direct numerical simulations (DNS) of centrifugal convection systems have conventionally employed periodic boundary conditions on the top/bottom surfaces, preventing the formation of rotation-dependent flow structures such as Ekman layers, Ekman pumping, and Stewartson layers. This geometric confinement, when combined with gravitational exclusion, restricts flow development to the plane orthogonal to the rotation axis, manifesting quasi-2D characteristics that may exhibit fundamental deviations from the practical experimental situation. Recent DNS investigations by \cite{2025_Direct_Yao} on the gravitational effect in centrifugal convection systems have demonstrated a clear flow regime transition: from centrifugal buoyancy-dominated horizontal flow organization to gravitational buoyancy-dominated vertical convection with increasing gravity. 

In CC experiments with strong rotation, the large Froude number (centrifugal force versus gravity) indicates a centrifugal-buoyancy-dominated system \citep{2025_Direct_Yao}. However, the flow often maintains three-dimensional characteristics rather than becoming purely two-dimensional. 
Numerous studies of rotating Rayleigh-B\'enard convection (RRBC) demonstrate that rotating systems develop distinct boundary layers at no-slip walls \citep{1953_The_Chandrasekhar, 1983_Stability_Lucas, 1996_Rapidly_Julien, 2002_Turbulent_Vorobieff, 2021_The_Kunnen, 2022_Turbulent_Ecke}:
- Ekman layers (perpendicular to the rotation axis)
- Stewartson layers (parallel to the rotation axis)
both play essential roles in heat transfer and flow structures.
In RRBC, rotation increases the critical Rayleigh number for convection onset and typically suppresses heat transport. However, when the Ekman layer ($\delta_E\sim Ek^{1/2}$) overlaps with the thermal boundary layer, Ekman pumping can enhance heat transfer in certain parameter ranges \citep{2009_Prandtl_Zhong, 2018_Effect_Chong, 2024_Critical_Anas}.
Strong zonal flows are observed at vertical sidewalls, along with Stewartson layers of thickness $\delta_{st}\sim Ek^{1/3}$ \citep{2011_The_Kunnen, 2013_The_Kunnen, 2021_Boundary_Zhang, 2020_Boundary_Zhang}. The vertical flow in the Stewartson layer is considered as a balance to the flow into Ekman boundary layers induced by the interior anticyclonic circulation, but the origins of the flows are not well addressed \citep{2013_The_Kunnen, 2021_Boundary_Zhang}.

In contrast to RRBC, the rotational axis of the CC system is perpendicular to the temperature gradient and centrifugal acceleration. Consequently, the delay of convection onset is not expected in this configuration \citep{2017_Onset_Pitz,2024_Effect_Zhong}; Ekman layers are supposed to be developed on the sidewalls, and Stewartson layers are supposed to be developed at the thermal boundaries, forming a distinct boundary layer architecture. Three fundamental questions arise: Does the Stewartson layer physically persist in CC? What determines its strength magnitude? How does it modify heat transfer efficiency and reorganize global flow patterns? Additionally, does gravity influence vertical flow dynamics within this layer? These questions form the central focus of this investigation.

This study employs three-dimensional numerical simulations of centrifugal convection (CC) with imposed gravity and no-slip top/bottom boundaries to systematically investigate the Stewartson layer in CC. The rest of the paper is organized as follows: the governing equations and setup for DNS are presented in Section \ref{sec2}, and the results of the sidewalls, gravity and rotation effect on the Stewartson layer are discussed in Section \ref{sec3} with theoretical explanations. Finally, the full paper will be summarized in Section \ref{sec4}.

\section{Numerical Model} \label{sec2}
\begin{figure}
    \centering
    \includegraphics[width=0.65\linewidth]{{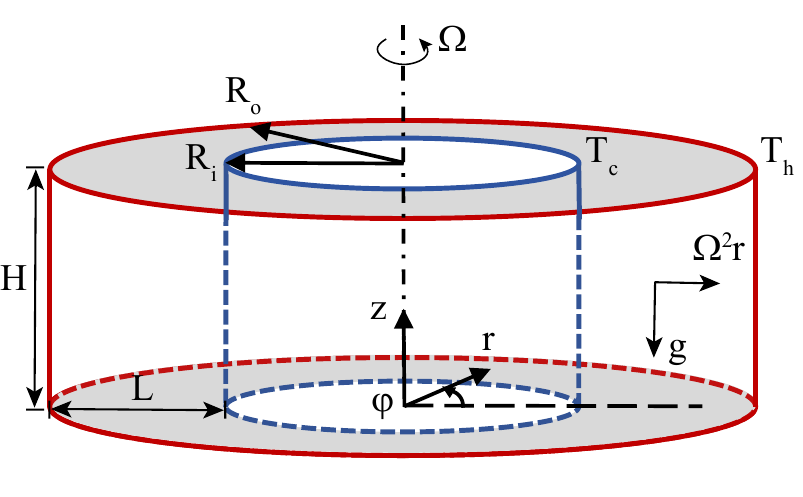}}
    \captionsetup{justification=justified}
    \caption{Schematic of the centrifugal convection system. The cylindrical annular chamber with inner and outer radii $R_{i,o}$ and height $H$ rotates about the vertical axis with angular velocity $\Omega$. The centrifugal acceleration $\Omega^2 r$ acts radially, while gravity $g$ acts axially, driving convection between the cold inner surface ($T_c$) and hot outer surface ($T_h$). No-slip boundary conditions are applied at all walls.}
	\label{fig: setup}
\end{figure}

The system schematic is shown in Figure \ref{fig: setup} using cylindrical coordinates $(r, \varphi, z)$. It consists of two coaxial cylinders rotating at the same angular velocity $\Omega$, with a cold inner cylinder ($T_c$) and hot outer cylinder ($T_h$). Key geometric parameters are the inner radius ($R_i$), outer radius ($R_o$), annular gap width $L=R_o-R_i$, and axial height $H$. The fluid's thermal expansion coefficient $\beta$, kinematic viscosity $\nu$, and thermal diffusivity $\kappa$ are treated as constant throughout the system.
The dimensionless governing equations following the Oberbeck-Boussinesq assumption in a rotating reference frame of angular velocity $\Omega$ read \citep{2020_Supergravitational_Jiang, 2022_Experimental_Jiang, 2023_From_Zhong}: 
\begin{equation}
    \bm{\nabla\cdot{u}}=0, \label{equ1}
\end{equation}
\begin{equation}
\frac{\partial T}{\partial t}+\bm{\nabla\cdot} (\bm{u}T)=\frac{1}{\sqrt{Ra_rPr}}\nabla^2T, \label{equ2}
\end{equation}
\begin{equation}
    \frac{\partial\bm{u}}{\partial t}+\bm{u\cdot\nabla u}=-\bm\nabla p-Ro^{-1}\bm{\hat{e_z}}\times\bm{u} +\sqrt{\frac{Pr}{Ra_r}}\nabla^2\bm{u}-\frac{2(1-\eta)}{(1+\eta)}rT\bm{\hat{e_r}}+\frac{Ra_g}{Ra_r}T\bm{\hat{e_z}}, \label{equ3}
\end{equation}
where $\bm{u}=(u,v,w)$, $T$, and $p$ represent the velocity, temperature, and pressure fields, respectively, with $\bm{\hat{e_r}}$ and $\bm{\hat{e_z}}$ being the unit vectors in the radial and axial directions. The quantities are nondimensionalized using the radial free-fall velocity $U=\sqrt{g_r\beta\Delta L}$, the gap width $L$, and the temperature difference $\Delta$, where $g_r=\Omega^2(R_i+R_o)/2$ represents the equivalent gravitational acceleration due to centrifugal rotation. The governing equations introduce four key dimensionless parameters: the centrifugal and gravitational Rayleigh numbers ($Ra_r$, $Ra_g$), the inverse Rossby number ($Ro^{-1}$), and the Prandtl number ($Pr$), defined as:
\begin{equation}
\begin{split}
    Ra_r=g_r\beta\Delta L^3/(\nu\kappa),~~ 
    Ra_g=g\beta\Delta L^3/(\nu\kappa)&,\\
    Ro^{-1}=2 \Omega L/U=2(\beta\Delta(R_o+R_i)/(2L))^{-1/2},~~ Pr=\nu/\kappa&.
 \end{split}   
\end{equation}

We introduce the inverse Froude number ($Fr^{-1}$) to characterize the relative strength of the gravitational acceleration and the centrifugal acceleration. It is defined as:
\begin{equation}
Fr^{-1} = g/g_r = Ra_g/Ra_r. 
\end{equation}\label{equ: Fr}
In this definition, gravity can be regarded as a disturbance to the centrifugal effect. When $Fr^{-1} \ll 1$ is satisfied, the influence of gravity on the main flow structure and the heat transfer can be neglected \citep{2025_Influence_liu}.
Additionally, in subsequent discussions, we will utilize the Ekman number ($Ek$) to facilitate comparison with previous studies. It is important to clarify that the Ekman number introduced herein serves solely to support the analysis of flow behaviors, rather than substituting for the originally defined control parameters. It is defined as follows:
\begin{equation}
   Ek=\nu/(\Omega H^2).
\end{equation}
In this system with aspect ratio $\Gamma=H/L$, it has the following relation with $Ra_r$, $Pr$ and $Ro^{-1}$:
\begin{equation}\label{equ: Ek}
   Ek=\sqrt{\frac{Pr}{Ra_r}}\frac{2}{Ro^{-1}}\Gamma^{-2}.
\end{equation}

In our simulations, the Prandtl number is fixed at $Pr=4.3$, alongside the geometry parameters, including the radius ratio $\eta=R_i/R_o=0.5$ and aspect ratio $\Gamma=H/L=1$. In particular, with $L=R_o-R_i$ as the characteristic length, the dimensionless radius of the inner cylinder and the dimensionless radius of the outer cylinder are
\begin{equation}
r_i =R_i/L=\eta/(1-\eta),
r_o =R_o/L=1/(1-\eta).
\end{equation}
In the present study, since $\eta$ is fixed at 0.5, the dimensionless inner and outer ring radii are 1 and 2, respectively. The no-slip and isothermal boundary conditions are applied for the inner and outer cylinders, whereas the no-slip and adiabatic boundary conditions are applied for the top and bottom covers. Considering the computational cost and the symmetry of convection vortices distribution, $1/4$ circle is adopted as the computational domain in all the cases  \citep{2022_Effects_Wang}. The details about the simulations can be found in Appendix \ref{App:1}.


\section{Results and Discussion} \label{sec3}
\subsection{Effect of sidewalls (top and bottom lids)}\label{sec3-1}

Based on previous studies, when gravity is neglected and the periodic boundary conditions are used for the top and bottom surfaces, the strong rotation tends to suppress vertical variation of the convection flow \citep{2020_Supergravitational_Jiang, 2022_Effects_Wang, 2023_From_Zhong}. This phenomenon can be attributed to the fact that the effect of the Coriolis force on the motion of a rotating fluid follows the Taylor-Proudman theorem, which results in a flow in a rotating vessel that is approximately quasi-two-dimensional. However, 
if there are top and bottom plates (lids) in the system, the presence of mechanical friction at the top and bottom walls will restrict the movement of the fluid, thus changing the pressure at the boundaries and creating a pressure difference with the body region. This results in the formation of vortices and will exert a significant influence on the flow structure \citep{1964_Wedemeyer_Unsteady,2002_Wind_Zhemin,2004_Introduction_Holton}.

\begin{figure}
    \centering
    \includegraphics[width=1\linewidth]{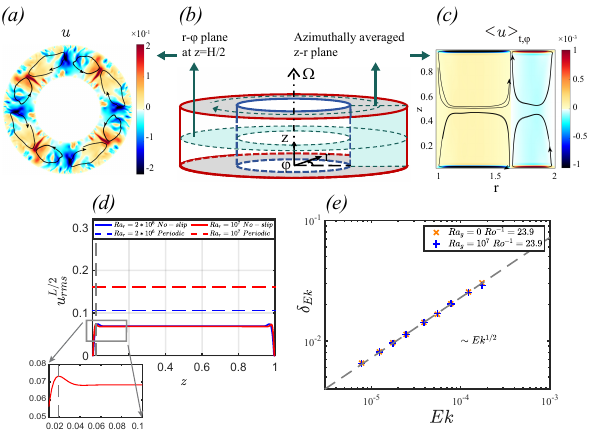}
 \captionsetup{justification=justified}
    \caption{(a) The contour indicates the instantaneous radial velocity component $u$ for $Ra_r = 10^8, Ra_g=0$ and $Ro^{-1} = 23.9$. The black curves in the figure denote the typical streamlines, and the arrows represent the flow direction. Note that a 1/4 circle is adopted as the computational domain here, and the complete circumference is drawn to match the schematic diagram. (b) The schematic diagram of the two types of planes selected in the system. (c) The contour indicates temporal and azimuthal averages of the radial velocity component $u$ for $Ra_r = 10^7, Ra_g=0$ and $Ro^{-1} = 23.9$. The black curves in the figure denote the typical streamlines, and the arrows represent the flow direction. (d) For $Ra_g=0$, the temporally and azimuthally averaged radial velocity r.m.s. profile in the vertical direction at $ r=1.5$. Dashed lines and solid lines of the same color represent the radial velocities under periodic boundary conditions and no-slip boundary conditions (applied to the top and bottom plates), respectively, for the same set of control parameters. The distance between the maximum radial velocity and the bottom plate is the thickness of the Ekman layer $\delta_{Ek}$ (gray dashed line). The small image in the lower-left corner shows the same data near the bottom plate. (e) The thickness of the Ekman layer $\delta_{Ek}$ varies with $Ek$.}
	\label{fig: Ekman}
\end{figure}

{Figure \ref{fig: Ekman}(a) displays the contour plot of the instantaneous radial velocity on the $r-\varphi$ plane with no-slip boundary conditions applied to the top and bottom plates. The methodology for selecting the plane is illustrated in figure \ref{fig: Ekman}(b), where the inner circular ring coincides with the inner cylinder and the outer circular ring corresponds to the outer cylinder. The black curves in the figure denote the typical streamlines of the convective vortices. Four pairs of convective vortices throughout the azimuthal direction are formed by the hot and cold plumes, which are detached from the thermal boundary layer and driven by centrifugal acceleration. Despite the effects of geometric curvature and Coriolis forces, the flow structure of convection remains highly similar to the classical RBC. The effect of the geometric curvature on the zonal flow and heat transfer in the CC system has been explored in detail in \cite{2022_Effects_Wang}. }

{Figure \ref{fig: Ekman}(c) presents the time- and azimuthally averaged radial velocity. The approach for plane selection is also demonstrated in figure \ref{fig: Ekman}(b), with the left-hand side of the panel (\(r = 1\)) coinciding with the inner cylinder and the right-hand side (\(r = 2\)) corresponding to the outer cylinder. It can be seen from the figure that with no-slip boundary conditions imposed on the top and bottom plates, the flow transitions from a quasi-two-dimensional state \citep{2020_Supergravitational_Jiang} to a three-dimensional regime due to the viscous effects of the top and bottom boundary, and the Ekman layer is observed. When the horizontal flows in the system meet the boundary, including the thermal plumes and zonal flow, the friction loss on the boundary leads to a vertical pressure gradient, and this Ekman pumping effect leads to vertical vortices \citep{Davidson_2013,John_Rotating_2012}. This secondary flow is also similar to the generation of the vertical vortex when a cup of tea is stirred \citep{2004_Introduction_Holton}. In our system, the streamlines in figure \ref{fig: Ekman}(c) show that four vortex structures are formed in the $r-z$ plane.}

{In figure \ref{fig: Ekman}(d), dashed lines and solid lines of the same color represent the root-mean-square radial velocities under periodic boundary conditions and no-slip boundary conditions (applied to the top and bottom plates), respectively, for the same set of control parameters. It can be observed that near the top and bottom walls, the radial velocity under no-slip boundary conditions is smaller than that under periodic boundary conditions. This confirms the influence of the boundary friction on the bulk velocity.} We define the thickness of the Ekman layer ($\delta_{Ek}$) as the distance between the time- and azimuthally root-mean-square-averaged radial velocity maximum and the wall near the bottom plate. Its thickness follows $\delta_{Ek}\sim Ek^{1/2}$, as shown in the figure \ref{fig: Ekman}(e), agreeing with the supposed effect of rotation.

\subsection{Effect of gravity}

\begin{figure}
    \centering
    \includegraphics[width=\linewidth]{{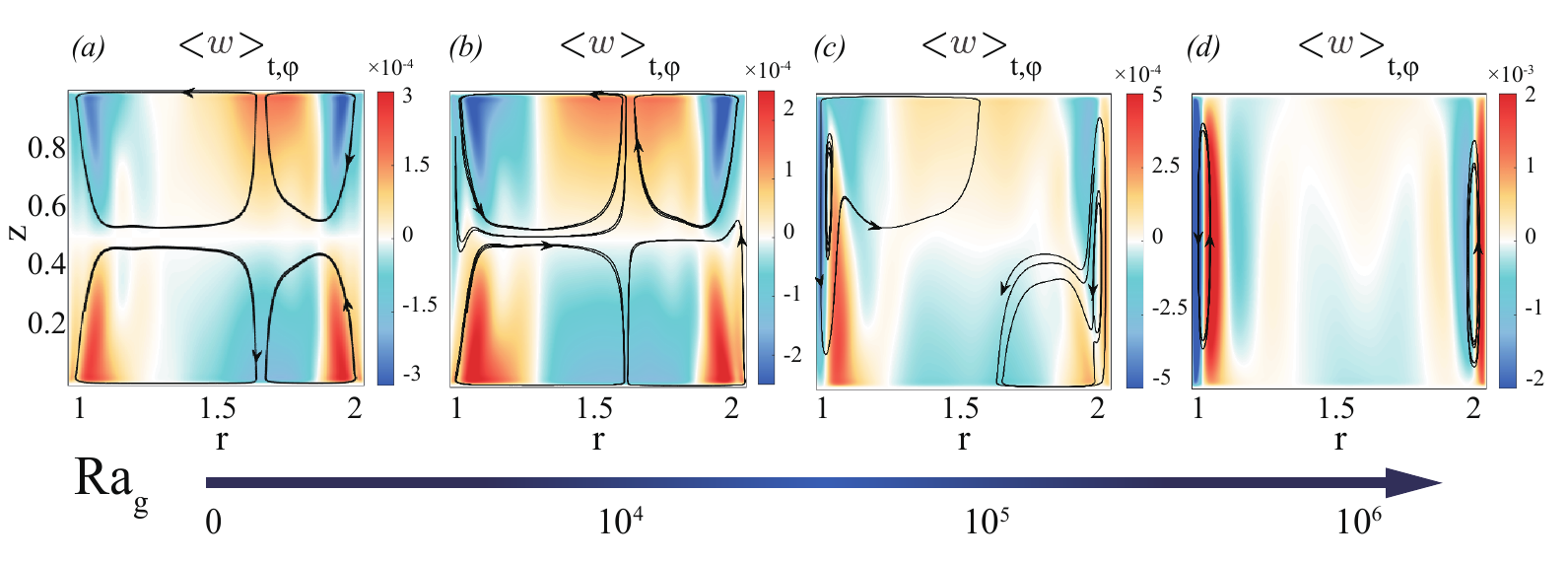}}
    \captionsetup{justification=justified}
     \caption{ 
    The contours indicate temporal and azimuthal averages of the velocity components $w$ for $Ra_r = 10^7$ and $Ro^{-1} = 23.9$. The black curves in the figure are representative streamlines, and the arrows represent the flow direction. The figures from left to right indicate the results for (a) $Ra_g = 0$, (b) $Ra_g = 10^4$, (c) $Ra_g = 10^5$, and (d) $Ra_g = 10^6$, respectively. Each panel is selected in the same manner shown in figure \ref{fig: Ekman}(b).
    }
	\label{fig: stewartson}
\end{figure}

To quantify the degree of flow three-dimensionalization under no-slip boundary conditions on the top and bottom walls, figure \ref{fig: stewartson} illustrates the four contour plots of time- and azimuthally averaged vertical velocity $\langle w\rangle_{t,\varphi}$ with increasing gravity strength at fixed $Ra_r$ and $Ro^{-1}$. Interestingly, as the imposed gravity increases, the vertical flow structure undergoes a significant transition, though its velocity magnitudes remain orders of magnitude weaker than the dominant horizontal convective motions.

When the cover plates exist, the secondary flow (the primary flow is the RBC rolls on the $r-\varphi$ plane) within the bulk region is roughly divided into four vortices of weak vertical velocity ($w\sim 10^{-4}$), which can be shown clearly by the streamlines in the figure \ref{fig: stewartson}(a). {The reasons for the formation of this flow structure have been discussed in Section \ref{sec3-1}}: the weak secondary flow structures arise from vertical pressure gradients induced by boundary-layer friction, where viscous dissipation at the confining walls modifies the momentum transport balance in near-wall regions. 
For systems with weak gravitational forcing ($Ra_g\le 10^4$), the main structure of the secondary flow remains consistent with the zero-gravity case, as shown in the figure \ref{fig: stewartson}(b). As the $Ra_g$ increases to $10^5$, the flow undergoes a symmetry-breaking transition, developing two elongated vortices close to the inner and outer walls. At higher $Ra_g$, these vortices merge into distinct shear layers near the sidewalls, called the Stewartson layer (figure \ref{fig: stewartson}(d))\citep{2011_The_Kunnen, 2022_Turbulent_Ecke}.

Interestingly, during the formation of the Stewartson layer, the system stays in centrifugal convection dominance as {$Fr^{-1}<1$} \citep{2025_Direct_Yao}. Compared to the main flow of convection, the secondary flow is weak but structurally clear. The Stewartson layer is formed and grows without significant changes of the main flow across different $Ra_g$, different with the formation of the Stewartson layer in the RRBC system, where the formation of the Stewartson layer is attributed to the bulk-driven circulation \citep{2011_The_Kunnen, 2013_The_Kunnen}. It can be inferred that the formation cause of the Stewartson layer in our system is more relevant to the gravity strength, and may not be identical to that observed in the RRBC system.

\begin{figure}
    \centering
    \includegraphics[width=\linewidth]{{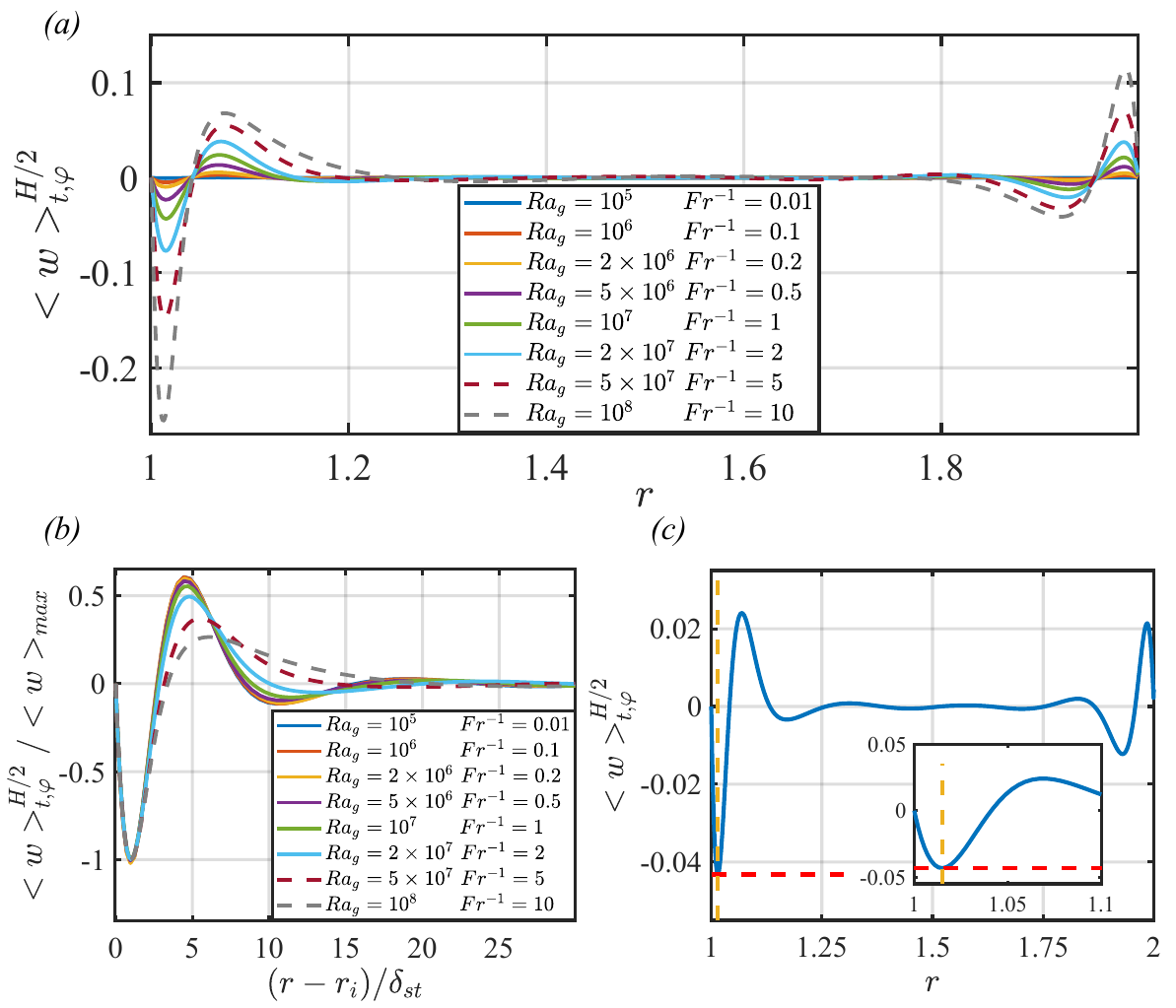}}
    \captionsetup{justification=justified}
    \caption{ {(a) The temporal and azimuthal averages of the velocity component $w$ in the radial direction at $z = H/2$ for $Ra_r = 10^7$ and $Ro^{-1} = 23.9$. When $Ra_g$ increases from $10^5$ to $10^8$, $Fr^{-1}$ gradually rises from 0.01 to 10. (b) The same data of vertical velocity for $1 \leq r \lesssim 1.5$ normalized by maximum velocity, and the radial coordinates are normalized by the thickness of the Stewartson layer. }(c) For $Ra_g = 10^7, Ra_r = 10^7$ and $Ro^{-1} = 23.9$, the time- and azimuthally averaged vertical velocity profile in radial direction at $z = H/2$; the amplitude $W_{st}$ is defined as the first peak of the absolute velocity maximum (red dashed line) and the thickness of the Stewartson layer $\delta_{st}$ is defined as the distance from the first peak to the closest boundary (yellow dashed line). The inset shows the same data near the inner cylinder.}
	\label{fig: fixRar}
\end{figure}

 To further investigate the characteristics and the formation of the Stewartson layer, we explore the influence of gravity on the configuration of the Stewartson layer at a fixed $Ra_r$, and the results are shown in the figure \ref{fig: fixRar}. The radial distribution of the averaged vertical velocity $w$ at $z=H/2$ is used to characterize the configuration of the Stewartson layer. The original profiles are displayed in the figure \ref{fig: fixRar}(a), and the normalized ones are displayed in the figure \ref{fig: fixRar}(b).
The radial distribution of mean vertical velocity exhibits consistent trends across varying $Ra_g$, with the velocity zero-crossing points aligning closely between cases. {Normalized velocity profiles further demonstrate that the shape of vertical velocity profiles is gravity-independent when the inverse Froude number is relatively small, as profiles collapse to a single curve. However, a clear divergence exists between the curves for $Fr^{-1} > 2$ and $Fr^{-1} \leq 2$.} 

{Previous studies have explored this transition process in detail \citep{2025_Influence_liu}. In the centrifugal convection (CC) system, the influence of the Froude number on the flow structure is mainly reflected in the evolution of the direction of large-scale circulation (LSC) and the vertical velocity. Gravity dominates at large $Fr^{-1}$, and a strong vertical LSC is formed, showing characteristics similar to vertical natural convection. When $0.2 < Fr^{-1} < 5$, the system is in a force balance region, centrifugal force weakens the vertical buoyancy driving, LSC shifts from vertical to horizontal direction, its intensity decreases accordingly, and the flow transitions to horizontal dominance. When $Fr^{-1}$ decreases further, centrifugal force dominates, axial flow is weak, horizontal LSC becomes dominant, Taylor columns are formed, and the system shows the characteristics of classical RBC.} 

{According to figure \ref{fig: fixRar}(b), we do not employ a strictly uniform interval division. Instead, the system is roughly divided into two regimes: the gravity-dominated regime ($Fr^{-1} > 2$, dashed lines) and the centrifugal-dominated regime ($Fr^{-1} \leq 2$, solid lines).} These results indicate that the mechanism of the Stewartson layer’s structure remains consistent in centrifugal-dominated regimes, where increased gravity solely modulates flow magnitude. However, as gravitational forcing transitions into dominance, the dynamics governing the layer’s configuration become progressively modified.

\begin{figure}
    \centering
    \includegraphics[width=\linewidth]{{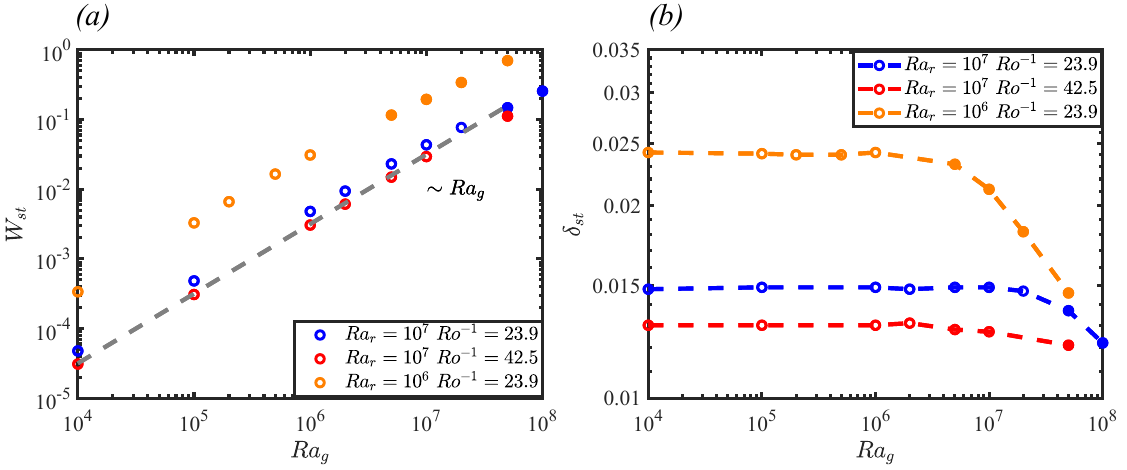}}
    \captionsetup{justification=justified}
    \caption{ {(a) The maximum velocity $W_{st}$ and (b) the thickness of Stewartson layer $\delta_{st}$ vary with $Ra_g$. The centrifugal and gravitational dominance cases are marked by the hollow and solid points, respectively.}}
	\label{fig: fixRarbls}
\end{figure}

Considering that the velocity zero-crossing points in figure \ref{fig: fixRar}(a) align closely between different cases, it appears that the thickness of the Stewartson layer does not change with $Ra_g$. To verify this observation, we need to quantify the thickness of the Stewartson layer. Referring to other articles about RRBC \citep{2013_The_Kunnen}, as shown in figure \ref{fig: fixRar}(c), we define the time- and azimuthally averaged vertical velocity maximum (absolute value) as the amplitude ($W_{st}$) and its distance to the wall as the thickness of the Stewartson layer ($\delta_{st}$). $W_{st}$ will be used to characterize the strength of the vertical velocity. Following these definitions, we further plot the velocity amplitude and the thickness of the Stewartson layer as a function of $Ra_g$ in figure \ref{fig: fixRarbls}. The centrifugal and gravitational dominance cases are marked by the hollow and solid points, respectively. Surprisingly, we can see that the vertical velocity amplitude in the Stewartson layer follows the scaling $W_{st}\sim Ra_g$ extremely well within the centrifugal dominance regime, among different parameter sets of $Ra_r$ and $Ro^{-1}$ from figure \ref{fig: fixRarbls}(a). Besides, as shown in figure \ref{fig: fixRarbls}(b), the thickness of the Stewartson layer remains nearly constant for {$Fr^{-1} \leq 2$}, consistent with our previous findings: the gravity controls the velocity amplitude and has no effect on the shape of the layer's profile. Meanwhile, some discrepancies in velocity amplitude and layer thickness emerge under gravity-dominated conditions, likely resulting from the shift in primary flow dynamics between Rayleigh-B\'enard and vertical convection regimes.

\subsection{Effect of rotation}

\begin{figure}
    \centering
    \includegraphics[width=\linewidth]{{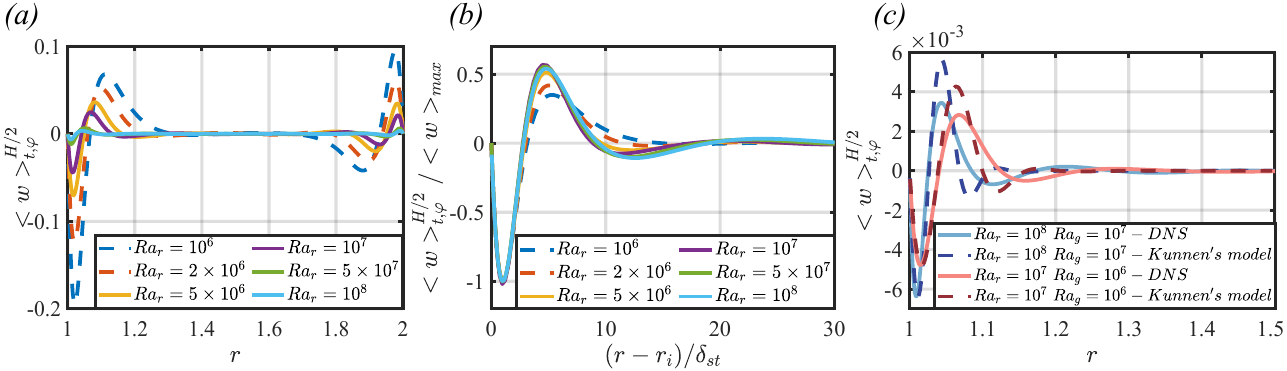}}
    \captionsetup{justification=justified}
    \caption{{
    (a) The temporal and azimuthal averages of the velocity components $w$ in the radial direction at $z = H/2$ for $Ra_g = 10^7$ and $Ro^{-1} = 23.9$. When $Ra_r$ increases from $10^6$ to $10^8$, $Fr$ gradually decreases from 10 to 0.1. (b) The same data of vertical velocity for $1 \leq r \lesssim 1.5$ normalized by maximum velocity, and the radial coordinates are normalized by the thickness of Stewartson layer. (c) Profiles of vertical velocity near the sidewall at height $z = H/2$, from DNS (solid lines) and theory (dashed lines). Theoretical formulae are derived from Kunnen's ideal model \citep{2013_The_Kunnen}.}}
	\label{fig: fixRag_vz} 
\end{figure}

Having examined the impact of gravity on the system, we now turn our attention to the effects of rotation on the Stewartson layer. In the RRBC system, the Stewartson boundary layer is found to consist of a sandwich structure of two boundary layers of typical thicknesses $Ek^{1/3}$ and $Ek^{1/4}$ \citep{2013_The_Kunnen}. According to the definition of Ekman number (the equation \ref{equ: Ek}), we maintain a constant value for $Ra_g$ while varying the $Ra_r$, and the effect of the inverse Rossby number $Ro^{-1}$ is also checked. 

Figure \ref{fig: fixRag_vz}(a) illustrates the radial distribution of the averaged vertical velocities, and the normalized profiles are presented in figure \ref{fig: fixRag_vz}(b). From figure\ref{fig: fixRag_vz}(a), the positions of the maximum velocities differ from each other, indicating that the $Ra_r$ affects both the thickness of the boundary layer and the strength of internal flow. Meanwhile, the shape of the profiles remains consistent in the centrifugal dominance regime {($Fr^{-1} \leq 2$)}, as shown in the normalized profiles. The dashed lines denote gravity-dominated flows {($Fr^{-1} > 2$)}. Their inner-layer profiles align closely with centrifugally dominant curves, whereas the outer layer exhibits marked deviation. This divergence likely arises from stronger coupling between the secondary flow and bulk flow dynamics.

In addition to this, we compared the results of the simulation with the theoretical model proposed by \cite{2013_The_Kunnen}. 
{Kunnen simplified the governing equations of the RRBC system and derived a theoretical solution of vertical velocity by introducing three key assumptions: first, viscous effects are negligible except in regions near the walls; second, a time-independent and azimuthally uniform mean circulation pattern exists; and third, the Rossby number ($Ro$) is negligibly small. With the velocity magnitude being the only fitted quantities, we apply Kunnen's theoretical solutions to our mean vertical velocity profiles at two different parameter sets, and the results are shown in the figure \ref{fig: fixRag_vz}(c).} Analytically derived profiles closely match simulated velocity distributions in near-wall regions. Mismatches emerge at radial positions farther from the sidewalls, possibly due to the influence of bulk flow. This finding also suggests that the dynamic velocity boundary layer of the secondary flow near the sidewall within our system is the same as the Stewartson layer observed in the RRBC system, and the local flow structure shares analogous characteristics.

\begin{figure}
    \centering
    \includegraphics[width=\linewidth]{{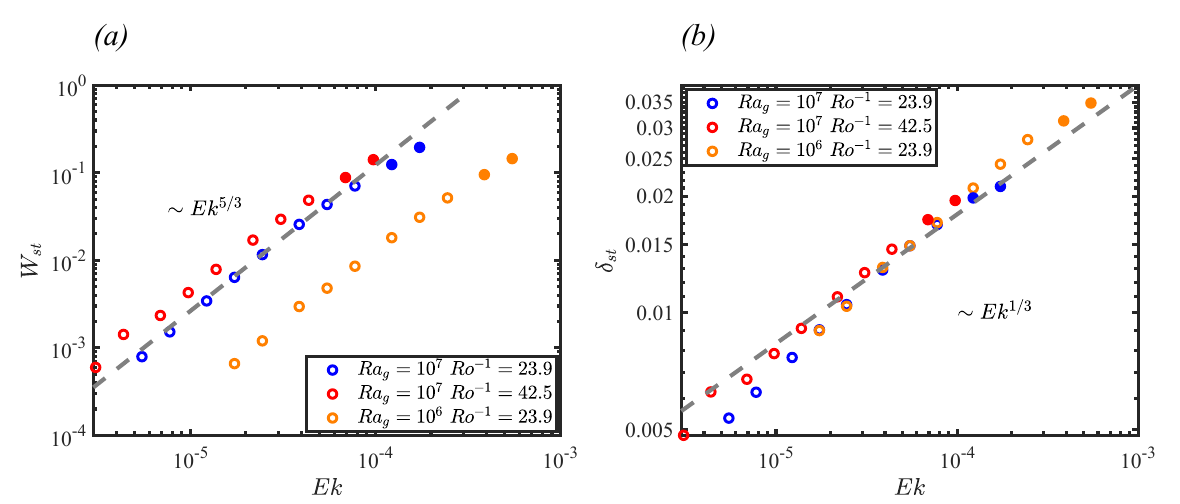}}
    \captionsetup{justification=justified}
    \caption{ (a) The maximum velocity $W_{st}$ and (b) the thickness of Stewartson layer $\delta_{st}$ varies with $Ek$. Cases under centrifugal dominance and gravitational dominance are marked as hollow and solid points, respectively. }
	\label{fig: fixRag_layer} 
\end{figure}

We further quantitatively analysed the effect of {rotation} on the features of the Stewartson layer, including the flow strength and the layer thickness defined by the first vertical velocity peak near the surface, and the results of different parameter set $\{Ra_g, Ro^{-1}\}$ are shown in the figure \ref{fig: fixRag_layer}. {It should be emphasized that when investigating the influence of rotation, the control parameter we use to characterize the rotational intensity is always $Ra_r$. $Ek$ is adopted as the abscissa in figure \ref{fig: fixRag_layer} for the assistance of the rotational physics and better comparisons with the relevant studies \citep{2013_The_Kunnen}.} For the thickness shown in figure \ref{fig: fixRag_layer}(b), all the data across varying $\{Ra_r, Ra_g, Ro^{-1}\}$ collapse onto the scaling $\delta_{st}\sim Ek^{1/3}$, demonstrating that the Ekman number exclusively governs the thickness of the inner Stewartson layer in our system. This result is also consistent with previous studies on rotating flow \citep{1993_Asymptotic_Herrmann, 2011_The_Kunnen, 2020_Boundary_Zhang, 2021_Boundary_Zhang}.
{In RRBC, there exists a unique wall mode with a typical two-layer structure near the sidewall, which is a concept under linear stability analysis, from which eigenfunctions of vertical velocity in the system can be derived \citep{2024_Zhang_Wall}. Both wall modes of RRBC and the Stewartson layer are rotation-regulated boundary layers with consistent scaling laws and similar vertical velocity radial profiles. Meanwhile, there are also a lot of differences between the CC and RRBC systems in temperature boundary conditions, dominant bulk flow, and gravitational influence. Though the different boundary conditions and gravity settings give rise to a wide diversity of flow phenomena, their core physical mechanisms may remain consistent.}
In addition, the outer layer with thickness $Ek^{1/4}$ is not observed in our simulation yet, which may be affected by the bulk flow dynamics. 

As for the flow strength, the figure \ref{fig: fixRag_layer}(a) shows that the velocity amplitude satisfies a scaling law of $W_{st}\sim Ek^{5/3}$, while it is also influenced by both $Ra_g$ and $Ro^{-1}$. A close scaling law follows $Ek^{1.8}$ is also observed in the previous study on the RRBC system \citep{2013_The_Kunnen}, and we will explain the mechanism behind this in the theoretical analysis section.

{In addition to the definition of boundary layer thickness adopted in this study (as shown in Figure 4(c) earlier, where $\delta_{st}$ is defined as the distance from the wall to the first peak of the time- and azimuthally-averaged vertical velocity), alternative definitions in the literature include using the distance from the wall to the local peak of the root-mean-square (r.m.s.) vertical velocity, and applying the slope method to determine the boundary layer thickness \citep{2010_Kunne_Experimental,2015_Gastine_Turbulent,2020_Boundary_Zhang}. For reference, the Appendix \ref{App:3} includes the boundary layer thickness values obtained via the "r.m.s. averaging + slope method" definition, as well as the variation of $W_{st}$ with the Ekman number.}

\subsection{Theoretical explanation}

Through previous analyses of the simulation results, we find that both rotational forcing ($Ek$) and vertical gravity ($Ra_g$) exert dominant control over the Stewartson layer dynamics. The Ekman number governs the configuration with an inner layer thickness $\delta_{st}\sim Ek^{1/3}$, while the combined influence of $Ek$ and $Ra_g$ governs the internal flow strength $W_{st}\sim Ra_gEk^{5/3}$. Whereas $Ek$’s role aligns with well-established scaling laws in rotating flows and RRBC systems, the explicit dependence on vertical gravity seems to emerge as a newly identified mechanism. In the RRBC system, the gravity exerts a global influence on large-scale convection, with its effects inherently hardly decoupling. However, in the CC system dominated by centrifugal force, the primary flow structure holds on the horizontal $r-\varphi$ plane, which offers a great opportunity to study the effect of $Ra_g$. 


Based on our findings, the Stewartson boundary layer is formed only when the vertical gravity exists and the flow strength is proportional to the gravitational intensity. Therefore, we have a new assumption: the internal vertical flow inside the Stewartson layer is induced by the vertical buoyancy, and the vertical velocity we measure in the mid-plane acts as a terminal velocity \citep{1990_stirring_griffiths, 2015_Cagney_Temperature}. Hence, there exists a force balance between viscosity and buoyancy in the vertical momentum equation insides the inner Stewartson layer, that is:
\begin{equation}
    \nu\nabla^2w^*\sim\beta g\delta T^*\label{equ4},
\end{equation}
where $w^*$ and $\delta T^*$ are dimensional vertical velocity and temperature difference. Considering the fact that in our simulations, the Stewartson layer is always inside the thermal boundary layer, we assume $\delta T^*$ is at the order of the global temperature difference $\Delta$. Use the same quantity scales as the governing equation \eqref{equ3}, then we have:
\begin{equation}
\sqrt{\frac{Pr}{Ra_r}}\nabla^2w\sim\frac{Ra_g}{Ra_r}.\label{equ5}
\end{equation}

As demonstrated in figures \ref{fig: fixRar} and \ref{fig: fixRag_vz}, the distribution of the normalized mean vertical velocity along the radial direction in the buoyancy dominant regime exhibits a similar configuration  
and $Ek^{1/3}$ is the radial characteristic scale. Hence, we take the maximum vertical velocity $W_{st}$, there is 
\begin{equation}
    \nabla^2w\sim Ek^{-2/3}W_{st}.\label{equ6}
\end{equation}

Equations \ref{equ5} and \ref{equ6} can be combined to obtain:
\begin{equation}
    W_{st}\sim (Ra_rPr)^{-1/2}Ek^{2/3}Ra_g.\label{equ7}
\end{equation}


Based on the relation between $Ek$ and $Ra_r$, $Ro^{-1}$ (shown in the equation \eqref{equ: Ek}), $Ra_r$ can be canceled and we finally get the scaling of $W_{st}$:
\begin{equation}
   {W_{st}\sim Ek^{5/3}Ro^{-1}Ra_gPr^{-1}\Gamma^2.}\label{equ9}
\end{equation}

\begin{figure}
    \centering
    \includegraphics[width=0.7\linewidth]{{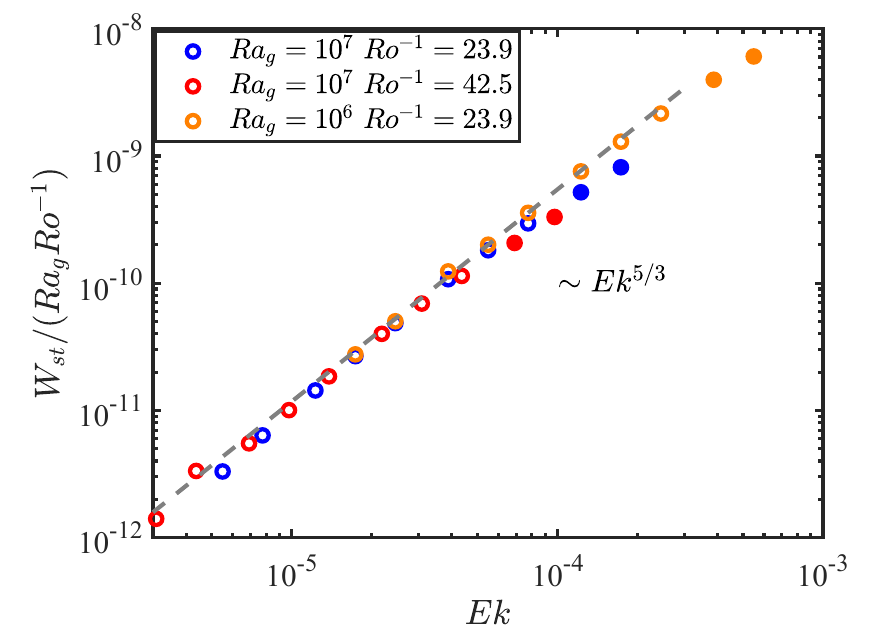}}
    \captionsetup{justification=justified}
    \caption{ The maximum velocity $W_{st}$ normalized by $Ra_gRo^{-1}$ varies with $Ek$. There is a clear scaling law $W_{st}/(Ra_gRo^{-1}) \sim Ek^{5/3}$. Cases under centrifugal dominance and gravitational dominance are marked as hollow and solid points, respectively. }
	\label{fig: Theory} 
\end{figure}

Surprisingly, both scaling laws $W_{st}\sim Ek^{5/3}$ and $W_{st}\sim Ra_g$ arise in the viscosity-buoyancy balance. Furthermore, following our theoretical results, we check the relation between normalized vertical velocity $W_{st}/(Ra_gRo^{-1})$ with $Ek$, and the results are presented in figure \ref{fig: Theory}. It is clear that all the data points collapse onto the single line with the predicted scaling, in perfect agreement with the derived equation \eqref{equ9}. 

As for the data points in the gravity-dominated regime (solid points), deviations from the theoretical curve are observed. When gravity fully dominates, the system turns to vertical convection. The thermal flow near the boundary layer is undergoing an accelerated process by buoyancy, and the vertical velocity gradually approaches the free-fall velocity:

\begin{equation}
    w\sim \sqrt{\beta\Delta gH}\sim Ra_g^{1/2}.\label{equ10}
\end{equation}
 
This result is consistent with the scaling law in laminar vertical convection \citep{2016_Momentum_Shishkina}, taking place when $Ra_g\gg Ra_r$ {($Fr^{-1}\gg 1$)}. Although our simulations have not touched that parameter region, it is still observed that the velocity deviates downwards from the predicted value, in agreement with the decreasing scaling exponent on $Ra_g$.

{}The theory provides a satisfactory explanation of the results found above about $W_{st}$. We recognize that the Stewartson boundary layer in CC is generated by gravitational buoyancy and shaped by the geostrophic balance of rotation.

\section{Conclusion} \label{sec4}

In this paper, we study the secondary flow structure in a more realistic CC system with the bottom and top plates (lids) and gravity.  With the boundary confinement, the Ekman layer is observed on the bottom and top surfaces, while the Stewartson layer is observed beyond a critical gravitational forcing. As gravity increases, the initial top-bottom symmetric quadrupolar vortex structure is broken, and two elongated vortex structures evolve near the inner and outer sidewalls and form the Stewartson layers. 
Furthermore, we performed a comprehensive analysis of the characteristics of the Stewartson layers, including the typical vertical velocity profiles, the internal flow strength, and the layer thickness $\delta_{st}$. It is found that the vertical velocity profiles hold the same shape after normalization, in line with Kunnen's theoretical solutions. Quantitative analyses confirm that the thickness of the Stewartson layer is governed by the rotation effect with $\delta_{st}\sim Ek^{1/3}$ and the flow strength is jointly modulated by both rotation and gravitational buoyancy through the composite scaling $W_{st}\sim Ra_gEk^{5/3}$. Moreover, as the gravitational buoyancy increases to dominance, the change of primary flow structure modifies the Stewartson boundary, inducing a departure from centrifugally dominated scaling laws.

Based on the findings, we demonstrate that the vertical flow inside the Stewartson layer is induced by the gravitational buoyancy and eventually reaches a terminal state with the balance between viscosity and buoyancy. Meanwhile, we establish that the length scale of the Stewartson layer is determined only by the Ekman number as $Ek^{1/3}$. From these judgments, we derived a theoretical model, which predicts that the vertical velocity amplitude satisfies $W_{st}\sim Ek^{5/3}Ro^{-1}Ra_gPr^{-1}$. There is a remarkably good agreement between the theory and the simulation results across varying $Ra_r$, $Ra_g$, and $Ro^{-1}$, confirming that in the CC system, the Stewartson layer is induced by the gravity and constrained by the geostrophic balance of rotation.

The present study primarily focuses on the vertical mean flow as the key indicator for characterizing the Stewartson layer. However, the gravitational and rotational effects on the other two velocity components may also play significant roles: the radial velocity is closely linked to convective heat transfer mechanisms, while the azimuthal velocity influences boundary layer morphology. Their mean flow and fluctuation statistics warrant detailed investigation in future studies. Nevertheless, the impact of the Stewartson layer on heat transport remains unclear. In Appendix \ref{App:2}, we present the variation of the Nusselt number ($Nu$) under different boundary conditions and gravitational Rayleigh numbers ($Ra_g$). No significant changes in heat transfer are observed within the current parameter range, suggesting the need for further investigation. Due to computational constraints, our simulations are limited to $Ra_r = 10^9$, which does not reach the ultimate regime. Future large-scale simulations at higher $Ra_r$ values—if computationally feasible—could help determine whether the Stewartson layer influences the viscous boundary layer transition and facilitates the onset of the ultimate regime.

\backsection [Acknowledgements] {We thank D. Lohse, and O. Shishkina for insightful discussions.}

\backsection [Funding]{This work was supported by NSFC Excellence Research Group Program for ‘Multiscale Problems in Nonlinear Mechanics’ (No. 12588201), and the New Cornerstone Science Foundation through the New Cornerstone Investigator Program and the XPLORER PRIZE.}

\backsection[Declaration of interests] {The authors report no conflict of interest.}

\appendix
\section{Numerical details}\label{App:1}
Our code is a further development of the open-source program AFiD for the classical RBC thermal convection system, where the effects of Coriolis force, gravity, and centrifugal buoyancy are considered in the new system. The meshing is done using a staggered grid, while the second-order central difference treats the spatial derivative terms, and the time-derivative terms are treated by the third-order stepwise Runge-Kutta method combined with the Crank-Nicholson implicit terms. The Courant-Friedrichs-Lewy (CFL) number is set to a maximum of 0.8 to ensure the temporal stability of the explicit format terms during integration \citep{1928_Uber_Courant}. In general, the flow state reaches statistical stability after more than 150 dimensionless times of simulation, followed by more than 350 dimensionless times of simulation for data collection for the sake of statistics. 

Previous studies have shown that the flow field in a CC system is periodic in the circumferential direction, and it is possible to halve the circumferential simulation region from a full circumference or even take only a $1/4$ circumference to reduce the computational effort. By ensuring that at least one pair of convective vortices is included in the computational domain, the Nu number and turbulence structure obtained will be similar to those obtained by taking the full circumference and will not affect the subsequent studies \citep{2022_Effects_Wang}.

Therefore, we also compared the Nusselt numbers and the thicknesses of the Stewartson layer, as well as the magnitudes calculated for full, 1/2, and 1/4 circles, as shown in the first three rows of Table \ref{table0} below. It can be seen that when changing the size of the selected circumference, the relative deviation of the results obtained for the $1/2$ and $1/4$ circle is negligible, around $3.5\%$, if the $Nu$, $\delta_{st}$, and $W_{st}$ calculated for the full circle are used as standard values. This is within the acceptable error range, showing that the simulation of partial circumferences along the circumference will not affect our subsequent research. What's more, we also verified the effect of radial and axial meshing on the results, displayed in Table \ref{table0}.

 \begin{table}
   \centering
   \caption{Grid resolution check}
   \label{table0}
   \resizebox{\textwidth}{!}{
     \begin{tabular}{cccccccccccccc}
     \hline
    $Ra_r$ & $Ra_g$ & $\Phi$ & $N_r 	\times N_{\varphi} \times N_z$ & $Ro^{-1}$ & $Nu$  & $\operatorname{Diff}_{Nu}$ & $W_{st}$ & $\delta_{st}$ \\
\hline
          $10^7$ & $10^7$ & $2\pi$ & $128\times1536\times768$ & 42.5  & 12.47  & 0.42\% & $2.84\times10^{-2}$ & $1.28\times10^{-2}$ \\
     $10^7$ & $10^7$ & $\pi$ & $192\times768\times384$ & 42.5  & 12.52  & 0.34\% & $2.92\times10^{-2}$ & $1.28\times10^{-2}$ \\
     $10^7$ & $10^7$ & $1/2\pi$ & $192\times384\times512$ & 42.5  & 12.47  & 0.67\% & $2.94\times10^{-2}$ & $1.27\times10^{-2}$ \\
     $10^7$ & $10^6$ & $1/2\pi$ & $192\times384\times192$ & 23.9  & 12.10  & 1.68\% & $4.49\times10^{-3}$ & $1.45\times10^{-2}$ \\
     $10^7$ & $10^6$ & $1/2\pi$ & $192\times384\times384$ & 23.9  & 12.03  & 0.41\% & $4.80\times10^{-3}$ & $1.49\times10^{-2}$ \\
     $10^7$ & $10^6$ & $1/2\pi$ & $192\times384\times512$ & 23.9  & 11.93  & 1.64\% & $4.63\times10^{-3}$ & $1.48\times10^{-2}$ \\
     $10^7$ & $10^6$ & $1/2\pi$ & $128\times384\times384$ & 23.9  & 12.02  & 0.17\% & $4.88\times10^{-3}$ & $1.48\times10^{-2}$ \\
     $10^7$ & $10^6$ & $1/2\pi$ & $144\times384\times384$ & 23.9  & 11.96  & 1.63\% & $4.80\times10^{-3}$ & $1.49\times10^{-2}$ \\
     \end{tabular}%
     }
   \label{tab:addlabel}%
 \end{table}%

The parameters of the main simulations considered in this work are listed in Table \ref{table1} and Table \ref{table2}, which are at $Ro^{-1}=23.9$ and $Ro^{-1}=42.5$, 
respectively. The columns in Table \ref{table1} and Table \ref{table2} from left to right indicate the centrifugal Rayleigh numbers $Ra_r$, the gravitational Rayleigh numbers $Ra_g$, the resolution in the radial, azimuthal and vertical direction $N_r \times N_{\varphi} \times N_z$, the Ekman numbers $Ek$, the Nusselt number of heat transfer $Nu$ and its relative difference of two halves $\operatorname{Diff}_{Nu}$, the velocity amplitude $W_{st}$, the thickness of the Stewartson layer $\delta_{st}$, the posterior check on the maximum grid spacing $\Delta_g$ by the global Kolmogorov length $\eta_K=(\nu^3/\varepsilon)^{1/4}$ and the number of grid points inside the Stewartson ($N_{St}$) and temperature boundary layers ($N_T$). Note that the properties of the Stewartson layer in the table denote the one on the inner cylinder. Moreover, the global mean kinetic energy dissipation rate $\varepsilon$ is estimated in the centrifugal dominance regimes ($Ra_r>Ra_g$) by the exact relations of CC: $\varepsilon=\nu^{3}L^{-4}f(\eta)Pr^{-2}Ra(Nu-1)$   \citep{2022_Spectra_Wang}, where the heat transfer is slightly affected by gravity, and the gravitational term in the energy balance can be ignored for the resolution check. $f(\eta)={2(\eta-1)}/{((1+\eta)ln(\eta))}$ is a correction factor.

 \begin{table}
   \centering
   \caption{{Numerical details for simulations at $Ro^{-1}=23.9$}}
   \label{table1}
    \resizebox{\textwidth}{!}{
     \begin{tabular}{cccccccccccccc}
     \hline
     $No.$ & $Ra_r$ & $Ra_g$ & $N_r 	\times N_{\varphi} \times N_z$ & $Ek$  & $Nu$  & $\operatorname{Diff}_{Nu}$ & $W_{st}$ & $\delta_{st}$ & $\Delta_g/\eta_k$ & $N_{St}$ & $N_T$ \\
     \hline
     1     & $10^5$ & $10^6$ & $144\times384\times432$ & $5.49\times10^{-4}$ & 4.89  & 1.01\% & $1.45\times10^{-1}$ & $3.48\times10^{-2}$ & $-$   & 12    & 23  \\
     2     & $2\times10^5$ & $10^6$ & $128\times256\times256$ & $3.88\times10^{-4}$ & 5.23  & 1.45\% & $9.51\times10^{-2}$ & $3.13\times10^{-2}$ & $-$   & 10    & 19  \\
     3     & $5\times10^5$ & $10^6$ & $128\times256\times256$ & $2.45\times10^{-4}$ & 5.58  & 0.41\% & $5.16\times10^{-2}$ & $2.80\times10^{-2}$ & $-$   & 9     & 20  \\
     4     & $10^6$ & 0     & $128\times384\times384$ & $1.74\times10^{-4}$ & 6.47  & 1.27\% & $-$   & $-$   & 0.256  & $-$   & 18  \\
     5     & $10^6$ & $10^4$ & $128\times384\times384$ & $1.74\times10^{-4}$ & 6.49  & 0.44\% & $3.37\times10^{-4}$ & $2.42\times10^{-2}$ & 0.256  & 8     & 18  \\
     6     & $10^6$ & $10^5$ & $128\times384\times384$ & $1.74\times10^{-4}$ & 6.42  & 0.21\% & $3.29\times10^{-3}$ & $2.41\times10^{-2}$ & 0.255  & 8     & 18  \\
     7     & $10^6$ & $2\times10^5$ & $128\times384\times384$ & $1.74\times10^{-4}$ & 6.50  & 1.49\% & $6.62\times10^{-3}$ & $2.40\times10^{-2}$ & 0.256  & 8     & 18  \\
     8     & $10^6$ & $5\times10^5$ & $128\times384\times384$ & $1.74\times10^{-4}$ & 6.59  & 0.37\% & $1.65\times10^{-2}$ & $2.40\times10^{-2}$ & 0.257  & 8     & 18  \\
     9     & $10^6$ & $10^6$ & $128\times384\times384$ & $1.74\times10^{-4}$ & 6.57  & 1.58\% & $3.10\times10^{-2}$ & $2.42\times10^{-2}$ & $-$   & 8     & 18  \\
     10    & $10^6$ & $5\times10^6$ & $128\times384\times384$ & $1.74\times10^{-4}$ & 8.18  & 0.43\% & $1.16\times10^{-1}$ & $2.32\times10^{-2}$ & $-$   & 8     & 15  \\
     11    & $10^6$ & $10^7$ & $144\times384\times432$ & $1.74\times10^{-4}$ & 10.27  & 0.12\% & $1.95\times10^{-1}$ & $2.12\times10^{-2}$ & $-$   & 8     & 14  \\
     12    & $10^6$ & $2\times10^7$ & $192\times384\times384$ & $1.74\times10^{-4}$ & 13.99  & 0.14\% & $3.41\times10^{-1}$ & $1.82\times10^{-2}$ & $-$   & 11    & 16  \\
     13    & $10^6$ & $5\times10^7$ & $192\times512\times512$ & $1.74\times10^{-4}$ & 20.36  & 0.03\% & $7.02\times10^{-1}$ & $1.46\times10^{-2}$ & $-$   & 9     & 12  \\
     14    & $2\times10^6$ & 0     & $128\times384\times384$ & $1.23\times10^{-4}$ & 7.62  & 0.26\% & $-$   & $-$   & 0.317  & $-$   & 16  \\
     15    & $2\times10^6$ & $10^6$ & $128\times384\times384$ & $1.23\times10^{-4}$ & 7.65  & 1.71\% & $1.81\times10^{-2}$ & $2.10\times10^{-2}$ & 0.317  & 7     & 16  \\
     16    & $2\times10^6$ & $10^7$ & $128\times384\times384$ & $1.23\times10^{-4}$ & 10.19  & 1.74\% & $1.24\times10^{-1}$ & $1.98\times10^{-2}$ & $-$   & 7     & 13  \\
     17    & $5\times10^6$ & 0     & $128\times384\times384$ & $7.76\times10^{-5}$ & 9.86  & 1.14\% & $-$   & $-$   & 0.425  & $-$   & 13  \\
     18    & $5\times10^6$ & $10^6$ & $144\times384\times384$ & $7.76\times10^{-5}$ & 9.85  & 1.00\% & $8.55\times10^{-3}$ & $1.71\times10^{-2}$ & 0.382  & 7     & 15  \\
     19    & $5\times10^6$ & $10^7$ & $192\times384\times384$ & $7.76\times10^{-5}$ & 11.25  & 0.48\% & $7.06\times10^{-2}$ & $1.69\times10^{-2}$ & $-$   & 10    & 19  \\
     20    & $10^7$ & 0     & $192\times512\times512$ & $5.49\times10^{-5}$ & 12.12  & 0.48\% & $-$   & $-$   & 0.369  & $-$   & 18  \\
     21    & $10^7$ & $10^4$ & $192\times384\times384$ & $5.49\times10^{-5}$ & 12.22  & 0.91\% & $4.77\times10^{-5}$ & $1.48\times10^{-2}$ & 0.411  & 9     & 19  \\
     22    & $10^7$ & $10^5$ & $192\times384\times384$ & $5.49\times10^{-5}$ & 11.94  & 1.56\% & $4.83\times10^{-4}$ & $1.49\times10^{-2}$ & 0.408  & 9     & 19  \\
     23    & $10^7$ & $10^6$ & $192\times384\times384$ & $5.49\times10^{-5}$ & 12.03  & 0.41\% & $4.80\times10^{-3}$ & $1.49\times10^{-2}$ & 0.368  & 9     & 19  \\
     24    & $10^7$ & $2\times10^6$ & $192\times384\times384$ & $5.49\times10^{-5}$ & 12.10  & 0.96\% & $9.40\times10^{-3}$ & $1.48\times10^{-2}$ & 0.410  & 9     & 19  \\
     25    & $10^7$ & $5\times10^6$ & $192\times384\times384$ & $5.49\times10^{-5}$ & 12.02  & 0.22\% & $2.31\times10^{-2}$ & $1.49\times10^{-2}$ & 0.409  & 9     & 19  \\
     26    & $10^7$ & $10^7$ & $192\times384\times384$ & $5.49\times10^{-5}$ & 12.47  & 1.03\% & $4.34\times10^{-2}$ & $1.49\times10^{-2}$ & $-$   & 9     & 18  \\
     27    & $10^7$ & $2\times10^7$ & $192\times384\times512$ & $5.49\times10^{-5}$ & 13.82  & 0.96\% & $7.71\times10^{-2}$ & $1.47\times10^{-2}$ & $-$   & 9     & 17  \\
     28    & $10^7$ & $5\times10^7$ & $240\times576\times576$ & $5.49\times10^{-5}$ & 17.16  & 0.88\% & $1.48\times10^{-1}$ & $1.37\times10^{-2}$ & $-$   & 12    & 19  \\
     29    & $10^7$ & $10^8$ & $288\times768\times768$ & $5.49\times10^{-5}$ & 21.84  & 1.07\% & $2.56\times10^{-1}$ & $1.22\times10^{-2}$ & $-$   & 14    & 20  \\
     30    & $2\times10^7$ & 0     & $192\times384\times512$ & $3.88\times10^{-5}$ & 14.54 & 0.26\% & $-$   & $-$   & 0.510  & $-$   & 17  \\
     31    & $2\times10^7$ & $10^6$ & $192\times432\times432$ & $3.88\times10^{-5}$ & 14.79  & 1.94\% & $2.96\times10^{-3}$ & $1.31\times10^{-2}$ & 0.461  & 9     & 16 \\
     32    & $2\times10^7$ & $10^7$ & $192\times384\times512$ & $3.88\times10^{-5}$ & 14.69  & 1.27\% & $2.57\times10^{-2}$ & $1.29\times10^{-2}$ & 0.460  & 8     & 16  \\
     33    & $5\times10^7$ & 0     & $240\times576\times576$ & $2.45\times10^{-5}$ & 19.21  & 1.13\% & $-$   & $-$   & 0.502  & $-$   & 18  \\
     34    & $5\times10^7$ & $10^6$ & $192\times432\times432$ & $2.45\times10^{-5}$ & 19.06  & 0.90\% & $1.20\times10^{-3}$ & $1.04\times10^{-2}$ & 0.617  & 7     & 14 \\
     35    & $5\times10^7$ & $10^7$ & $240\times576\times576$ & $2.45\times10^{-5}$ & 18.76  & 1.72\% & $1.18\times10^{-2}$ & $1.05\times10^{-2}$ & 0.500  & 10    & 18  \\
     36    & $10^8$ & 0     & $288\times768\times768$ & $1.74\times10^{-5}$ & 22.70  & 0.29\% & $-$   & $-$   & 0.527  & $-$   & 20  \\
     37    & $10^8$ & $10^6$ & $288\times768\times768$ & $1.74\times10^{-5}$ & 22.67  & 0.11\% & $6.59\times10^{-4}$ & $9.00\times10^{-3}$ & 0.527  & 11    & 20 \\
     38    & $10^8$ & $10^7$ & $288\times768\times768$ & $1.74\times10^{-5}$ & 22.89  & 1.45\% & $6.38\times10^{-3}$ & $9.04\times10^{-3}$ & 0.527  & 11    & 20  \\
     39    & $2\times10^8$ & 0     & $288\times768\times768$ & $1.23\times10^{-5}$ & 28.40  & 0.71\% & $-$   & $-$   & 0.663  & $-$   & 17 \\
     40    & $2\times10^8$ & $10^7$ & $288\times768\times768$ & $1.23\times10^{-5}$ & 27.97  & 1.00\% & $3.43\times10^{-3}$ & $7.67\times10^{-3}$ & 0.660  & 10    & 17 \\
     41    & $5\times10^8$ & 0     & $288\times768\times768$ & $7.76\times10^{-6}$ & 37.02  & 1.73\% & $-$   & $-$   & 0.891  & $-$   & 15 \\
     42    & $5\times10^8$ & $10^7$ & $288\times768\times768$ & $7.76\times10^{-6}$ & 36.47  & 0.03\% & $1.52\times10^{-3}$ & $6.24\times10^{-3}$ & 0.888  & 9     & 15 \\
     43    & $10^9$ & $10^7$ & $512\times1152\times1152$ & $5.49\times10^{-6}$ & 44.07  & 1.53\% & $7.91\times10^{-4}$ & $5.35\times10^{-3}$ & 0.642  & 18    & 26 \\
     \end{tabular}%
     }
   \label{tab:addlabel}%
 \end{table}%

 \begin{table}
   \centering
   \caption{Numerical details for simulations at $Ro^{-1}=42.5$}
   \label{table2}
   \resizebox{\textwidth}{!}{
     \begin{tabular}{cccccccccccccc}
     \hline
    $No.$ & $Ra_r$ & $Ra_g$ & $N_r 	\times N_{\varphi} \times N_z$ & $Ek$  & $Nu$  & $\operatorname{Diff}_{Nu}$ & $W_{st}$ & $\delta_{st}$ & $\Delta_g/\eta_k$ & $N_{St}$ & $N_T$ \\
\hline
     1     & $10^6$ & $10^7$ & $144\times384\times432$ & $9.76\times10^{-5}$ & 8.95  & 0.60\% & $1.41\times10^{-1}$ & $1.95\times10^{-2}$ & $-$   & 8     & 16  \\
     2     & $2\times10^6$ & $10^7$ & $192\times384\times384$ & $6.90\times10^{-5}$ & 9.59  & 0.06\% & $8.81\times10^{-2}$ & $1.74\times10^{-2}$ & $-$   & 11    & 22  \\
     3     & $5\times10^6$ & $10^7$ & $192\times384\times384$ & $4.36\times10^{-5}$ & 10.92  & 0.84\% & $4.85\times10^{-2}$ & $1.46\times10^{-2}$ & $-$   & 9     & 20  \\
     4     & $10^7$ & $10^4$ & $192\times384\times512$ & $3.09\times10^{-5}$ & 11.99  & 1.46\% & $3.09\times10^{-5}$ & $1.30\times10^{-2}$ & 0.409  & 9     & 19  \\
     5     & $10^7$ & $10^5$ & $192\times384\times512$ & $3.09\times10^{-5}$ & 11.87  & 0.92\% & $3.08\times10^{-4}$ & $1.30\times10^{-2}$ & 0.408  & 9     & 19  \\
     6     & $10^7$ & $10^6$ & $192\times384\times512$ & $3.09\times10^{-5}$ & 11.92  & 0.02\% & $3.06\times10^{-3}$ & $1.30\times10^{-2}$ & 0.408  & 9     & 19  \\
     7     & $10^7$ & $2\times10^6$ & $192\times384\times512$ & $3.09\times10^{-5}$ & 11.96  & 1.71\% & $6.11\times10^{-3}$ & $1.31\times10^{-2}$ & 0.409  & 9     & 19  \\
     8     & $10^7$ & $5\times10^6$ & $192\times384\times512$ & $3.09\times10^{-5}$ & 12.21  & 0.27\% & $1.49\times10^{-2}$ & $1.28\times10^{-2}$ & 0.411  & 8     & 18  \\
     9     & $10^7$ & $10^7$ & $192\times384\times512$ & $3.09\times10^{-5}$ & 12.47  & 0.67\% & $2.94\times10^{-2}$ & $1.27\times10^{-2}$ & $-$   & 8     & 18  \\
     10    & $10^7$ & $5\times10^7$ & $240\times576\times576$ & $3.09\times10^{-5}$ & 16.31  & 1.29\% & $1.12\times10^{-1}$ & $1.21\times10^{-2}$ & $-$   & 11    & 20  \\
     11    & $2\times10^7$ & $10^7$ & $192\times384\times512$ & $2.18\times10^{-5}$ & 14.93  & 1.38\% & $1.70\times10^{-2}$ & $1.10\times10^{-2}$ & 0.462  & 8     & 16  \\
     12    & $5\times10^7$ & $10^7$ & $240\times576\times576$ & $1.38\times10^{-5}$ & 19.01  & 1.18\% & $7.87\times10^{-3}$ & $9.12\times10^{-3}$ & 0.503  & 9     & 18  \\
     13    & $10^8$ & $10^7$ & $288\times768\times768$ & $9.76\times10^{-6}$ & 23.39  & 1.01\% & $4.27\times10^{-3}$ & $7.85\times10^{-3}$ & 0.531  & 10    & 20 \\
     14    & $2\times10^8$ & $10^7$ & $288\times768\times768$ & $6.90\times10^{-6}$ & 28.51  & 1.63\% & $2.34\times10^{-3}$ & $6.74\times10^{-3}$ & 0.664  & 9     & 17 \\
     15    & $5\times10^8$ & $10^7$ & $288\times768\times768$ & $4.36\times10^{-6}$ & 37.30  & 1.41\% & $1.42\times10^{-3}$ & $6.25\times10^{-3}$ & 0.893  & 9     & 15 \\
     16    & $10^9$ & $10^7$ & $512\times1152\times1152$ & $3.09\times10^{-6}$ & 44.93  & 1.66\% & $5.97\times10^{-4}$ & $4.83\times10^{-3}$ & 0.645  & 15    & 26 \\
     \end{tabular}%
     }
   \label{tab:addlabel}%
 \end{table}%

\section{Effects on heat transfer}\label{App:2}

\begin{figure}
    \centering
    \includegraphics[width=0.6\linewidth]{{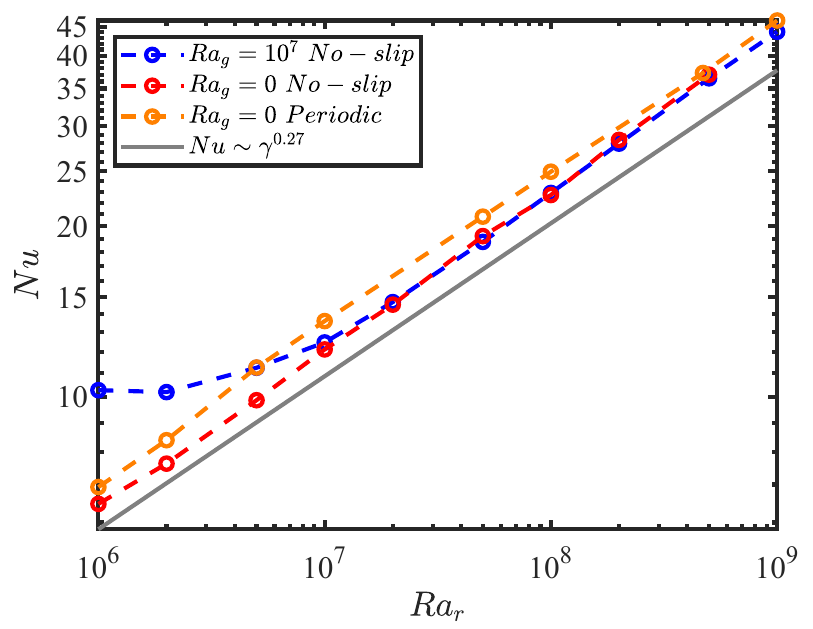}}
    \captionsetup{justification=justified}
    \caption{ The variation of $Nu$ under different boundary conditions at $Ro^{-1}=23.9$ with $Ra_r$.}
	\label{fig: Nu}
\end{figure}

Figure \ref{fig: Nu} shows the variation of $Nu$ under different boundary conditions and different $Ra_g$. Based on the previous discussion of the flow structure, when $Ra_g = 0$ and periodic boundary conditions are applied to the top and bottom walls (yellow points), there is no Ekman layer or Stewartson boundary layer. Keeping $Ra_g = 0$, when no-slip boundary conditions are applied to the top and bottom walls (red points), the Ekman layer can be observed near the walls, while when $Ra_g$ is large (blue points), the Stewartson layer appears. By comparing the $Nu$ values in these three cases, we can make some inferences about the effects of the Ekman layer and the Stewartson layer on heat transfer; however, no significant changes in heat transfer have been observed in the current parameter regime. Furthermore, the energy losses caused by friction at the wall surfaces and the influence of $Ra_g$ cannot be ruled out, which further complicates the process of drawing reliable conclusions. The influence of the velocity boundary layer, especially the Stewartson layer, on heat transfer requires further in-depth research. 

\section{Calculations of $W_{st}$ and $\delta_{st}$ using r.m.s. vertical velocity}\label{App:3}

\begin{figure}
    \centering
    \includegraphics[width=1\linewidth]{{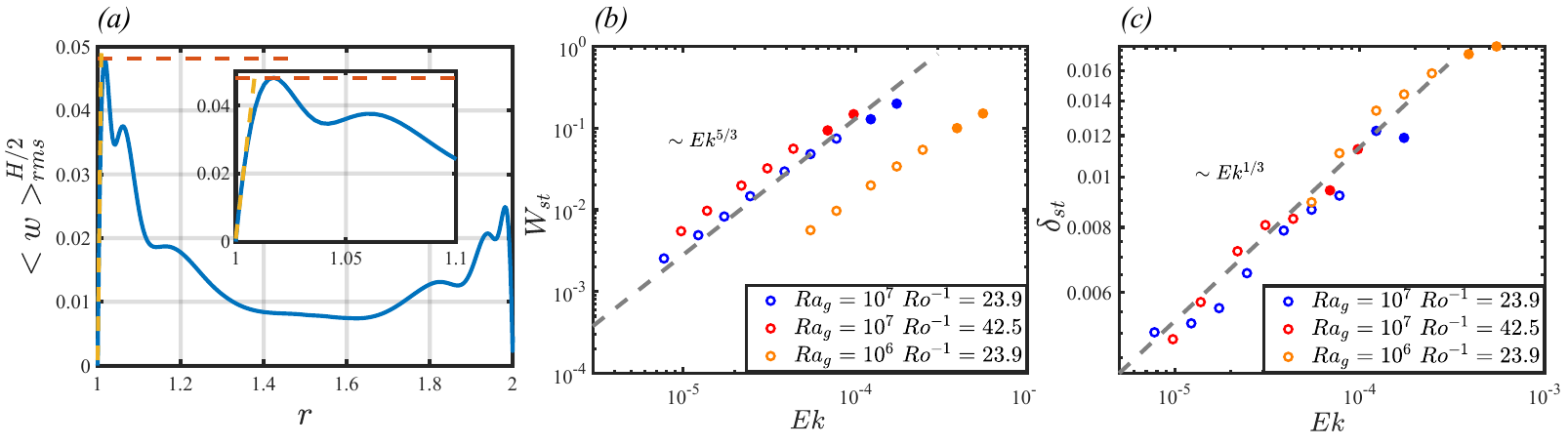}}
    \captionsetup{justification=justified}
    \caption{{(a) For $Ra_g=Rar=10^7,Ro^{-1}=23.9$, the root-mean-square vertical velocity profile in the radial direction at middle height $ z=H/2$. The yellow dashed line represents a linear fit performed on the  r.m.s. vertical velocity profile near the wall. A horizontal line (red dashed line) is drawn through the local maximum ($W_{st}$) of the $w$. The radial distance from the intersection of these two lines to the wall is defined as the Stewartson boundary layer thickness ($\delta_{st}$). Besides, the zoomed image in the upper-right corner shows the same data near the inner cylinder. (b) The maximum velocity $W_{st}$ and (c) the thickness of the Stewartson layer $\delta_{st}$ vary with $Ek$. Cases under centrifugal dominance and gravitational dominance are marked as hollow and solid points, respectively. }}
	\label{fig: st_msq}
\end{figure}

{Following another definition of boundary layer thickness by the r.m.s. profile \citep{2015_Gastine_Turbulent,2010_Kunne_Experimental,2020_Boundary_Zhang,2024_Zhang_Wall}, as shown in figure \ref{fig: st_msq}(a), we could define the thickness of the Stewartson boundary layer based on the intersection of tangents to the velocity profile. This definition draws on the "slope method" concept for thermal boundary layers \citep{1999_Verzicco_Prandtl}: first, a linear fit is performed on the r.m.s.  vertical velocity profile near the wall, then a horizontal line is drawn through the local maximum ($W_{st}$) of the vertical velocity, and the radial distance from the intersection of these two lines to the wall is defined as the Stewartson boundary layer thickness ($\delta_{st}$).}

{Under this definition, the relationships between $W_{st}$, $\delta_{st}$ and the Ekman number can be observed in figures \ref{fig: st_msq}(b) and (c). It is evident that $\delta_{st}$ remains agreeing well with the $\delta_{st}\sim Ek^{1/3}$ scaling law under these circumstances, as well as the strength of the internal flow $W_{st}\sim Ek^{5/3}$. 
These results sufficiently demonstrate that our conclusions are robust to the choices of the Stewartson boundary layer thickness definition.}

\bibliographystyle{jfm}
\bibliography{jfm}

\end{document}